\documentclass[twocolumn,english,aps]{revtex4}
\usepackage[T1]{fontenc}
\usepackage[latin9]{inputenc}
\usepackage{color}
\usepackage{amsmath}
\usepackage{graphicx}
\usepackage{esint}

\makeatletter

\newcommand{\noun}[1]{\textsc{#1}}

\@ifundefined{textcolor}{}
{%
 \definecolor{BLACK}{gray}{0}
 \definecolor{WHITE}{gray}{1}
 \definecolor{RED}{rgb}{1,0,0}
 \definecolor{GREEN}{rgb}{0,1,0}
 \definecolor{BLUE}{rgb}{0,0,1}
 \definecolor{CYAN}{cmyk}{1,0,0,0}
 \definecolor{MAGENTA}{cmyk}{0,1,0,0}
 \definecolor{YELLOW}{cmyk}{0,0,1,0}
 }

\renewcommand\[{\begin{equation}} 

\renewcommand\]{\end{equation}} 

\makeatother

\usepackage{babel}

\begin{document}

\title{Capillary instability in concentric-cylindrical shell: numerical
simulation and applications in composite microstructured fibers }

\author{D. S. Deng$^{\text{1, 2}}$, J.-C. Nave$^{\text{3}}$, X. Liang$^{\text{1, 3}}$,
S. G. Johnson$^{\text{1, 3}*}$, Y. Fink$^{\text{1, 2*}}$ }

\address{$^{\text{1}}$Research Laboratory of Electronics, $^{\text{2}}$Department
of Materials Science and Engineering, $^{\text{3}}$Department of
Mathematics, Massachusetts Institute of Technology, 77 Massachusetts
Avenue, Cambridge, Massachusetts 02139, USA $^{\text{*}}$ stevenj@mit.edu;
yoel@mit.edu.}
\begin{abstract}
Recent experimental observations have demonstrated interesting instability
phenomenon during thermal drawing of microstructured glass/polymer
fibers, and these observations motivate us to examine surface-tension-driven
instabilities in concentric cylindrical shells of viscous fluids.
In this paper, we focus on a single instability mechanism: classical
capillary instabilities in the form of radial fluctuations, solving
the full Navier--Stokes equations numerically. In equal-viscosity
cases where an analytical linear theory is available, we compare to
the full numerical solution and delineate the regime in which the
linear theory is valid. We also consider unequal-viscosity situations
(similar to experiments) in which there is no published linear theory,
and explain the numerical results with a simple asymptotic analysis.
These results are then applied to experimental thermal drawing systems.
We show that the observed instabilities are consistent with radial-fluctuation
analysis, but cannot be predicted by radial fluctuations alone---an
additional mechanism is required. We show how radial fluctuations
alone, however, can be used to analyze various candidate material
systems for thermal drawing, clearly ruling out some possibilities
while suggesting others that have not yet been considered in experiments.
\end{abstract}
\maketitle

\section{Introduction\label{sec:-Introduction}}

The classical capillary instability, the breakup of a cylindrical
liquid thread into a series of droplets, is perhaps one of the most
ubiquitous fluid instabilities and appears in a host of daily phenomena
\cite{Eggers1997,deGennes2002} from glass-wine tearing $ $\textcolor{blue}{{}
}\textcolor{black}{\cite{Cabazat92}}, and faucet dripping to ink-jet
printing. The study of capillary instability has a long history. In
1849, Plateau attributed the mechanism to surface tension: the breakup
process reduces the surface energy \cite{Plateau1849}. Lord Rayleigh
pioneered the application of linear stability analysis to quantitatively
characterize the growth rate at the onset of instability, and found
that a small disturbance is magnified exponentially with time \cite{Rayleigh1878}.
Subsequently, Tomotika investigated the effect of viscosity of the
surrounding fluid, showing that it acts as a deterrent to slow down
the instability growth rate \cite{Tomotika1935}. \textcolor{black}{Stone
and Brenner investigated instabilities in a two-fluid cylindrical
shell geometry with equal viscosities \cite{Stone1996}. }Many additional
phenomena have been investigated, such as the cascade structure in
a drop falling from a faucet \cite{Shi1994}, steady capillary jets
of sub-micrometer diameters \cite{Gannan2007}, and double-cone neck
shapes in nanojets \cite{Moseler2000}. Capillary instability also
offers a means of controlling and synthesizing diverse morphological
configurations. Examples include: a long chain of nanospheres generated
from tens-of-nanometer diameter wires at the melt state \cite{Molares2004,Karim2006};
polymer nanorods formed by annealing polymer nanotubes above the glass
transition point \cite{Chen2007}; and nanoparticle chains encapsulated
in nanotubes generated by reduction of nanowires at a sufficiently
high temperature \cite{Qin2008}. Moreover, instabilities of fluid
jets have numerous chemical and biological applications \cite{Stone2004,Squires2005}.
(An entirely different instability mechanism has attracted recent
interest in elastic or visco-elastic media, in which thin sheets under
tension form wrinkles driven by elastic instabilities \cite{Huang2002,Cerda2002,Cerda2003}.)

\textcolor{black}{Recently, a new class of glass and polymer microstructured
fibers, which are characterized by an embedded geometry of concentric
cylindrical shells, has emerged for a variety of applications in optics
\cite{Abouraddy2007,Deng2008,Deng2009}. Uniform-thickness cylindrical
shells (made of glass materials such as $\mathrm{\mathrm{As_{2}Se_{3}}}$
and $\mathrm{As_{2}S_{3}}$), down to sub-micrometer or even nanometer
thicknesses, have been successfully fabricated in glass materials,
by a drawing process whereby a large-scale {}``preform'' is heated
and pulled into a long thread, as depicted in Fig. \ref{fig:fiberdrawing}.
On the other hand, as the shell thickness is further reduced towards
the nanoscale, the thin cylindrical shell (made of the glass material
selenium, $\mathrm{Se}$), is observed to break up into an ordered
array of filaments; that is, the breakup of the cylindrical shell
occurs along the azimuthal direction while the axial continuity remains
intact \cite{Deng2008,Deng2009}.}\textcolor{blue}{{} }

Fiber drawing of cylindrical shells and other microstructured geometries
clearly opens up rich new areas for fluid instabilities and other
phenomena, and it is necessary to understand these phenomena in order
to determine what structures are attainable in drawn fibers. Although
it is a striking example, the observed azimuthal breakup process appears
to be a complicated matter---because surface tension does not produce
azimuthal instability in cylinders \cite{Chandrasekhar1961,Eggers2008}
or cylindrical shells \cite{Liang2010}, the azimuthal breakup must
be driven by the fiber draw-down process and/or by additional physics
such as visco-elastic effects or thermal gradients. Simulating the
entire draw-down process directly is very challenging, however, because
the lengthscales vary by $5$ orders of magnitude from the preform
($\mathrm{cm}$) to the drawn layers (sub-$\mathrm{\mu m}$). Another
potentially important breakup process, one that is amenable to study
even in the simplified case of a cylindrical-shell geometry, is axial
instability. Not only does the possibility of axial breakup impose
some limits on the practical materials for fiber drawing, but understanding
such breakup is arguably a prerequisite to understanding the azimuthal
breakup process, for two reasons. First, the azimuthal breakup process
produces cylindrical filaments, and it is important to understand
why these filaments do not exhibit further breakup into droplets (or
under what circumstances this should occur). Second, it is possible
that the draw-down process or other effects might couple fluctuations
in the axial and azimuthal directions, so understanding the timescales
of the axial breakup process is necessary as a first step in evaluating
whether it plays any physical role in driving other instabilities. 

Therefore, as a first step towards understanding the various instability
phenomena in drawn microstructured (cylindrical shell) fibers, we
investigate the impact of a single mechanism: c\textcolor{black}{lassical
Plateau--Rayleigh instability leading to radial fluctuations in cylindrical
fluids, here including the previously unstudied case of concentric
shells with different viscosity. To isolate this mechanism from other
forms of instability, we consider only cylindrically symmetrical geometries,
which also greatly simplifies the problem into two dimensions $(r,z)$.
(Indeed, this model is a satisfactory description of the cylindrical
filaments.) We model this situation by direct numerical simulation
via a finite-element method for the Navier--Stokes (NS) equations.
After first reproducing the known results from the linear theory for
equal viscosity, we are able to explore the regime beyond the limits
of the linear theory, which we show is accurate up to fluctuations
of about $10\%$ in radius. For the case of unequal-viscosity shells,}\textcolor{blue}{{}
}where there is not as yet a linear-theory analysis, \textcolor{black}{we
show that the instability timescale interpolates smoothly }between
two limits that can be understood by dimensional analysis. \textcolor{black}{Finally,
we apply our results to the experimental fiber-drawing situation,
where we obtain a necessary (but not sufficient) condition for stability
that can be used to guide the materials selection and the design of
the fabrication process by excluding certain materials combinations
from consideration. Our stability criterion is shown to be consistent
with the experimental observations. We also find that the stability
of the resulting filaments is consistent with the Rayleigh-Tomotika
model.}

\textcolor{black}{Motivated by the desire to understand the range
of attainable structures and improve their performance, previous theoretical
study of microstructured fiber drawing has considered a variety of
situations different from the one considered here. Air-hole deformation
and collapse was explored by numerical analysis of the continuous
drawing process of microstructured optical fibers \cite{Fitt2001,Xue2006}.
Surface-tension effects have been studied for their role in determining
the surface smoothness and the resulting optical loss of the final
fibers \cite{Robert2005}. The modeling of non-circular fibers has
also been investigated in order to design fiber-draw process for unusual
fiber shapes with a square or rectangular cross-section \cite{Griffiths2008}. }

This paper is organized as follows. We provide more background on
microstructured fibers in section~\ref{sec:Feature-Size-in}, and
review the governing equations and dimensionless groups pertaining
to fiber thermal-drawing processing in section~\ref{sec:Governing-Equations}.
Section~\ref{sec:-Simulation-Algorithm} describes the numerical
finite-element approach to solving the Navier--Stokes equations, and
section~V presents our simulation results for cylindrical shell.
$ $\textcolor{black}{{} Section~VI presents a radial stability map
to guide materials selection, which is established by linear-theory
calculations dependent on the shell radius, thickness, and viscosities.
In section~\ref{sec:Applications-in-Microstructured}, we discuss
the applications of capillary instability to our microstructured fibers
for viscous materials selection and limits on the ultimate feature
sizes. In particular, section~\ref{sub:azimuthal-preference} gives
a simple geometrical argument to explain why any instability mechanism
in fiber drawing will tend to favor azimuthal breakup (into filaments)
over axial breakup (into rings or droplets).}

\section{Feature Size in Composite Microstructured Fibers\label{sec:Feature-Size-in}}

\begin{figure}
\includegraphics[width=1\columnwidth]{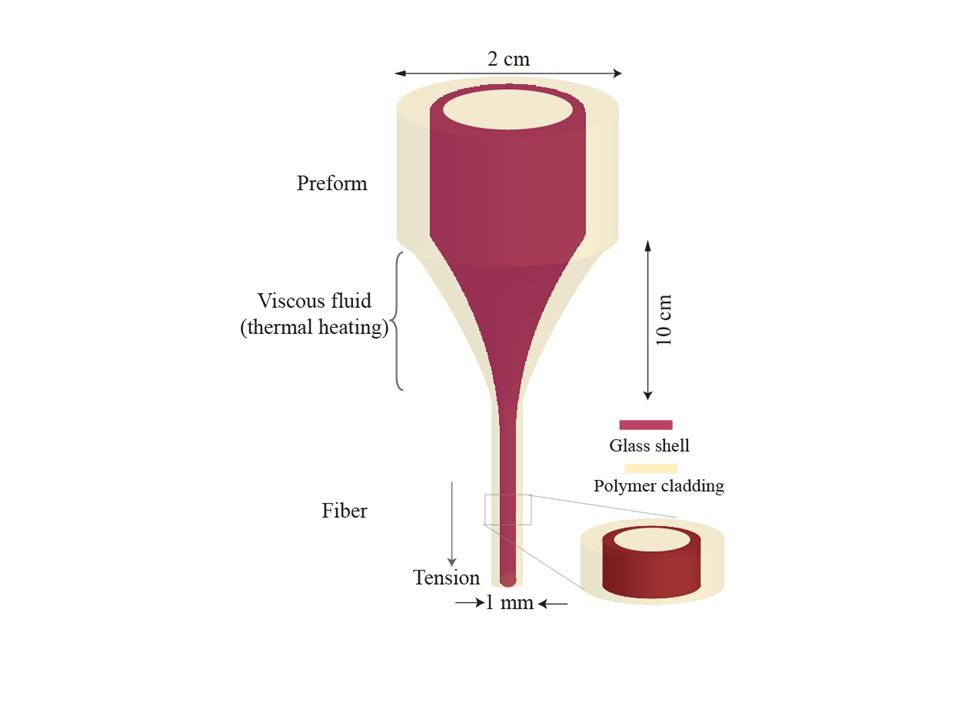}\caption{\label{fig:fiberdrawing}Optical-fiber thermal drawing. Preform is
heated at elevated temperature to viscous fluid, and stretched into
extended fibers by applied tension. This preform is specially designed
with a thin cylindrical shell in polymer matrix. }

\end{figure}

Most optical fibers are mainly made of a single material, silica glass.
Recent work, however, has generalized optical fiber manufacturing
to include microstructured fibers that combine multiple distinct materials
including metals, semiconductors, and insulators, which expand fiber-device
functionalities while retaining the simplicity of the thermal-drawing
fabrication approach \textcolor{black}{\cite{Abouraddy2007}}. For
example, a periodic cylindrical-shell multilayer structure has been
incorporated into a fiber to guide light in a hollow core with significantly
reduced loss for \textcolor{black}{laser surgery \cite{Temelkuran2002}.
The basic element of such a fiber is a cylindrical shell of one material
(a chalcogenide glass) surrounded by a {}``cladding'' of another
material (a thermoplastic polymer), as depicted in Fig. \ref{fig:fiberdrawing}.}
The fabrication process has four main steps to create this geometry.
(i) A glass film is thermally evaporated onto a polymer substrate.
(ii) The glass/polymer bi-layer film is tightly wrapped around a polymer
core. (iii) Additional layers of protective polymer cladding are then
rolled around the structure. (iv) The resulting centimeter-diameter
preform is fused into a single solid structure by heating under vacuum.
The solid preform is then heated into a viscous state and stretched
into an extended millimeter-diameter fiber by the application of axial
tension, as shown in Fig.\ref{fig:fiberdrawing}. 

\begin{figure}
\includegraphics[width=1\columnwidth]{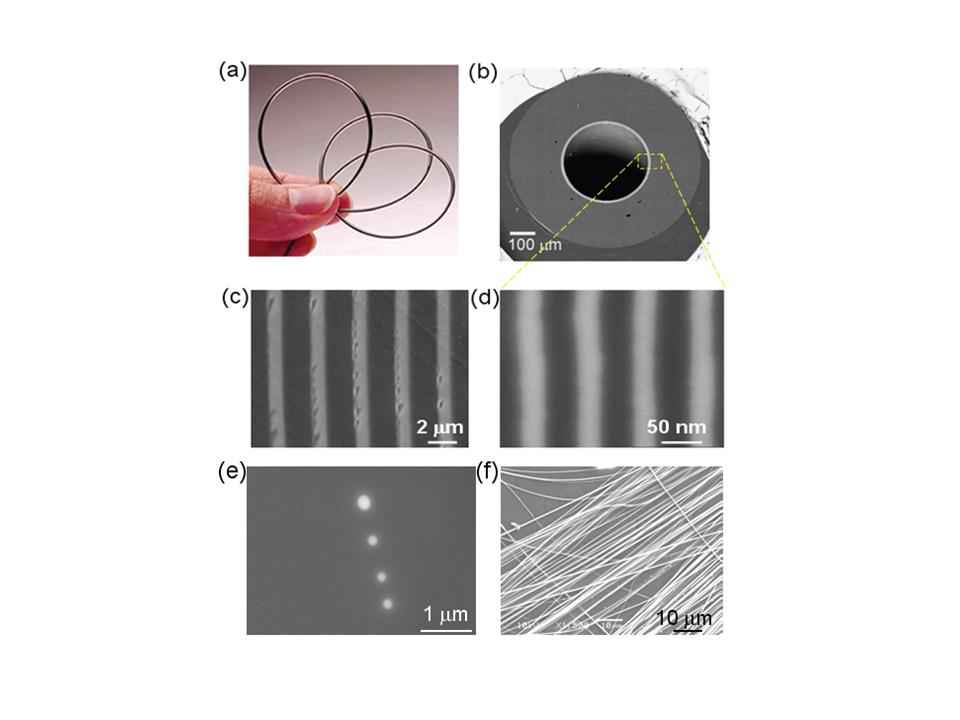}

\caption{\label{fig:SEM}SEM micrographs of cylindrical shells in fiber. (a)
Photograph of fiber. (b) SEM of fiber cross-section. Magnified view
of multilayer structures reveals the thickness of micrometer (c) and
tens of nanometers (d), respectively. Bright and dark color for glass
and polymer in SEM, respectively. \textcolor{black}{(e) showing layer
breakup into circles in the fiber cross-section, while (f) presenting
the continuous filaments obtained from fiber after dissolving polymer
matrix. }}

\end{figure}

\textcolor{black}{Uniform layer thicknesses, down to micrometers or
even to nanometers,} have been successfully achieved in fibers by
this method. \textcolor{black}{Mechanically flexible fibers with a
uniform diameter have been produced, as shown in Fig. \ref{fig:SEM}(a).
Fig. \ref{fig:SEM}(b) shows a typical Scanning Electron Microscope
(SEM) micrograph of a cross-section with $1$ $\mathrm{mm}$ fiber
diameter. Magnified SEM micrographs in Fig. \ref{fig:SEM}(c) and
(d), for two different fibers, reveal high-quality multiple-layer
structures with thicknesses on the order of micrometers and tens of
nanometers, respectively. }

\textcolor{black}{Ideally, the cross-section of the resulting fibers
retains the same structure and relative sizes of the components as
in the preform. If the layer thickness becomes too small, however,
we observed the layers to break up azimuthally, as shown in the cross-section
of Fig. \ref{fig:SEM}(e), while continuity along the axial direction
remains intact. In this fashion, a thin cylindrical shell breaks into
filament arrays embedded in a fiber \cite{Deng2008,Deng2009}. After
dissolving the polymer cladding, the resulting separated glass filaments
are shown in Fig. \ref{fig:SEM}(f) \cite{Deng2008}. }

\textcolor{black}{As a first step towards understanding these observations,
we will consider simplified geometries consisting of a thin cylindrical
shell in the cladding matrix. The question, here, is whether the classical
capillary instability from radial fluctuations is relevant to the
attained thin uniform shells and the observed azimuthal breakup at
further reduced thicknesses. Moreover, we want to know whether classical
capillary instability can provide guidance for materials selections
and fabrication processing in our microstructured fibers. }

\section{Governing Equations\label{sec:Governing-Equations}}

During thermal drawing, the temperature is set above the softening
points of all the materials, which consequently are in the viscous
fluid state to enable polymer and glass codrawing. To describe this
fluid flow, we consider the incompressible Navier--Stokes (NS) equations
\cite{Batchelor00}: \begin{equation}
\begin{cases}
\rho\left[\partial_{t}\vec{u}+\left(\vec{u}\cdot\nabla\right)\vec{u}\right]=-\nabla p+\nabla\cdot\left[\eta\frac{\nabla\vec{u}+\left(\nabla\vec{u}\right)^{T}}{2}\right]-\gamma\kappa\vec{n}\delta,\\
\nabla\cdot\vec{u}=0,\end{cases}\label{eq:NS}\end{equation}
where $ $ $\vec{u}$ is velocity, $p$ is pressure, $\rho$ is materials
density, $\eta$ is viscosity, $\gamma$ is interfacial tension between
glass and polymer, $\delta$ is the delta function ($\delta=1$$ $
at the polymer and glass interface and $\delta=0$ otherwise), $\kappa$
is curvature of interface, and $\vec{n}$ is a unit vector at interface. 

To identify the operating regime of the fiber drawing process, we
consider the relevant dimensionless numbers. The Reynolds number ($\mathrm{Re}$),
Froude number ($\mathrm{Fr}$), and capillary number($\mathrm{Ca}$)
are:

\begin{equation}
\mathrm{Re=\frac{\rho Uh}{\eta},Fr=\frac{U^{2}}{gh},Ca=\frac{\eta U}{\gamma},}\label{eq:Dimensionless}\end{equation}
 \textcolor{black}{where $\rho\approx10^{3}$ }$\mathrm{kg/m^{3}}$\textcolor{black}{{}
$ $ is the density of the materials, $g\approx10$ $\mathrm{m/s^{2}}$
is gravity, $U\approx5$ $\mathrm{mm/s}$ is drawing speed, $\eta\approx10^{5}$
$\mathrm{Pa\cdot s}$ is viscosity, $h\approx100$ $\mathrm{nm}$
is the layer thickness, and $\gamma=0.1$ $ $} $\mathrm{N/m}$\textcolor{black}{{}
$ $ is surface tension between polymer and glass \cite{Hart2004,Hartthesis}.
}Therefore, these dimensionless numbers in a typical fiber draw are
$\mathrm{Re}\approx10^{-10}$, $\mathrm{Fr}\approx10^{2}$ and $\mathrm{Ca}\approx10^{4}$.
\textcolor{black}{Small $\mathrm{Re}$ number, large $\mathrm{Fr}$
number, and large $\mathrm{Ca}$ number imply a weak inertia term,
negligible gravity, and dominant viscosity effects, respectively.
In addition,} \textcolor{black}{since the fiber diameter is $D\approx1$}$\mathrm{\mbox{}\mbox{}mm}$
and the length of the neck-down region is $L\approx10$ $\mathrm{cm}$,
the ratio \textcolor{black}{$D/L\approx1/100$} is much less than
$1$, and thus the complicated profile of neck-down cone is simplified
into a cylindrical shape for the purpose of easier analysis.

\section{Simulation Algorithm \label{sec:-Simulation-Algorithm}}

To develop a quantitative understanding of capillary instability in
a cylindrical-shell geometry, direct numerical simulation is performed
using the finite element method. In order to isolate the effect of
radial fluctuations, we impose cylindrical symmetry, so the numerical
simulation simplifies into a $\mathrm{2D}$ problem in the $(r,z)$
plane (where the $\nabla$ operations are replaced by their cylindrical
forms). \textcolor{black}{Although the low $\mathrm{Re}$ number means
that one could accurately neglect the $(\vec{u}\cdot\nabla)\vec{u}$
nonlinear inertia term and solve only the time-dependent Stokes problem,
we solve the full NS equations because of the convenience of using
available software supporting these equations and also to retain the
flexibility to consider large $\mathrm{Re}$ regimes in the future. }

\begin{figure}
\includegraphics[width=1\columnwidth]{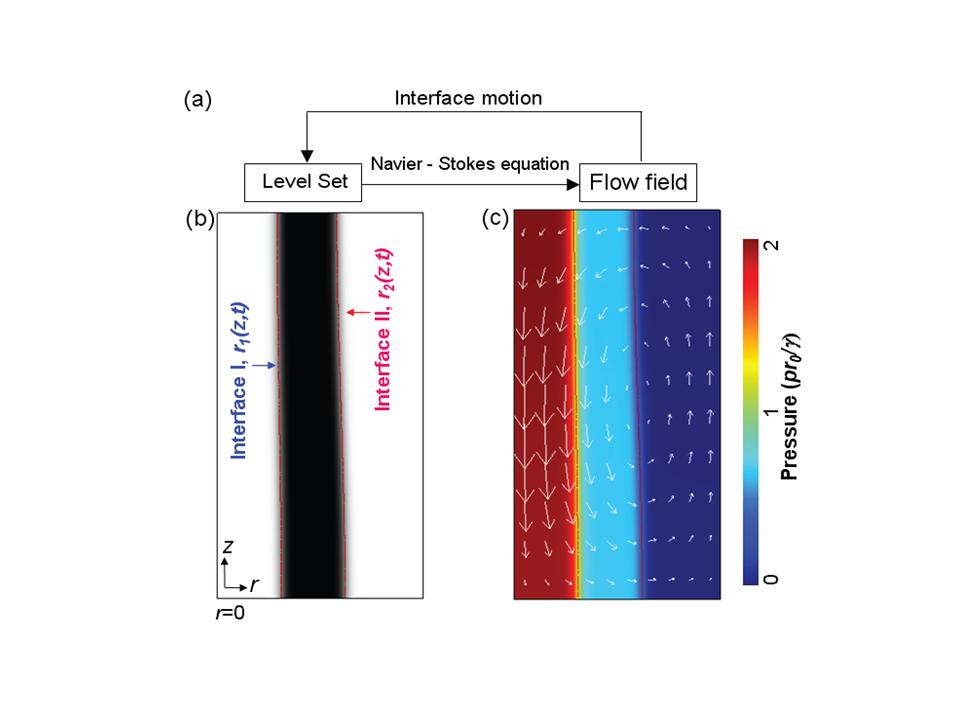}

\caption{\label{fig:Simulation-algorithm }Simulation algorithm. (a) Schematic
of algorithm. (b) Interfaces \textit{\textcolor{black}{$\mathrm{I}$}}
and $\mathrm{II}$ of cylindrical shell are defined by a level set
function. (c) Flow field is determined by the NS equations (color
scale for pressure and arrows for fluid velocity). }

\end{figure}

Our simulation algorithm is briefly presented in Fig. \ref{fig:Simulation-algorithm }(a).
The geometry of a cylindrical shell is defined by two interfaces $\mathrm{I}$
and $\mathrm{II}$ in Fig. \ref{fig:Simulation-algorithm }(b). The
flow field ($\vec{u},p$) is calculated from Navier--Stokes equations,
as seen in Fig. \ref{fig:Simulation-algorithm }(c). Consequently,
this new flow field generates interface motion, resulting in an updated
interface and an updated flow field. By these numerical iterations,
the interface gradually evolves with time. 

\textcolor{black}{More specifically, the time-dependent interfaces
$\mathrm{I}$ and $\mathrm{II}$ can be expressed as follows:}

\textcolor{black}{\begin{equation}
\begin{cases}
r_{1}(z,t)= & r_{0}+\varepsilon_{1}(z,t),\\
r_{2}(z,t)= & 2r_{0}+\varepsilon_{2}(z,t),\end{cases}\label{eq:radius}\end{equation}
where $r_{0},2r_{0}$ are the unperturbed radii of interface $\mathrm{I}$
and $\mathrm{II}$, $\varepsilon_{1}(z,t),\varepsilon_{2}(z,t)$ denote
the interfacial perturbations that grow with time. }

\textcolor{black}{The initial perturbations at interface $\mathrm{I}$
and $\mathrm{II}$ are given by cosine-wave shapes: }

\textcolor{black}{\begin{equation}
\begin{cases}
\varepsilon_{1}(z,0)= & \varepsilon_{0}\mathrm{cos}(2\pi z/\lambda),\\
\varepsilon_{2}(z,0)= & 2\varepsilon_{0}\mathrm{cos}(2\pi z/\lambda),\end{cases}\label{eq:intialcondition}\end{equation}
where $\varepsilon_{0},2\varepsilon_{0}$ are the perturbation amplitudes,
and $\lambda$ is the initial perturbation wavelength. }

\textcolor{black}{Numerical challenges in the simulations arise from
the nonlinearity, moving interfaces, interface singularities, and
the complex curvature \cite{Eggers1997,Scardovelli1999}}\textcolor{blue}{.
}A level-set\textcolor{black}{{} $ $ function} $\phi(\vec{x},t)$ is
coupled with the NS equations to track the interface (see Appendix
1) $\mbox{}$\textcolor{black}{{} \cite{Osher1988,Osher2001,Sethian2003}},
where the interface is located at the $\phi=0.5$ contour and the
$\phi$ evolution is given by $\vec{u}$ via:

\begin{equation}
\phi_{t}+\vec{u}\cdot\vec{\nabla}\phi=0.\label{eq:Levelset}\end{equation}
The local curvature ($\kappa$) at an interface is given in terms
of $\phi$ by: 

\begin{multline}
\kappa=\nabla\cdot\vec{n}=\nabla\cdot\frac{\nabla\phi}{|\nabla\phi|}\\
=\frac{(\phi_{r}^{2}\phi_{zz}+\phi_{z}^{2}\phi_{rr})+(\phi_{r}^{2}+\phi_{z}^{2})\phi_{r}/r-\phi_{r}\phi_{z}(\phi_{rz}+\phi_{zr})}{(\phi_{r}^{2}+\phi_{z}^{2})^{3/2}}.\label{eq:Curvature-1}\end{multline}
For numerical stability reasons, we add an artificial diffusion term
proportional to a small parameter $D$ in Eq. \ref{eq:Levelset} (see
Appendix 4); the convergence as $D\rightarrow0$ is discussed in the
next section. 

\section{\textcolor{black}{Simulation Results\label{sec:Simulation-Results}}}

\textcolor{black}{In the low $\mathrm{Re}$ regime, the NS equations
are linear with respect to the flow velocity $\vec{u}$ because the
inertia term} $(\vec{u}\cdot\nabla)\vec{u}$\textcolor{black}{{} $ $
is negligible, but they are not linear with respect to the geometry
of the interface shape except in the limit of small perturbations.
A linear theory for small geometric perturbations was formulated for
the cylinder by Tomotika \cite{Tomotika1935} and for a cylindrical
shell with a cladding of equal viscosity by Stone and Brenner \cite{Stone1996}.
Our full simulation should, of course, reproduce the results of the
linear theory for small perturbations, but also allows us to investigate
larger perturbations beyond the regime of the linear theory, and we
can also consider the more general case of a cylindrical shell with
a cladding of unequal viscosity.}

\subsection{Numerical convergence }

\begin{figure}
\includegraphics[width=1\columnwidth]{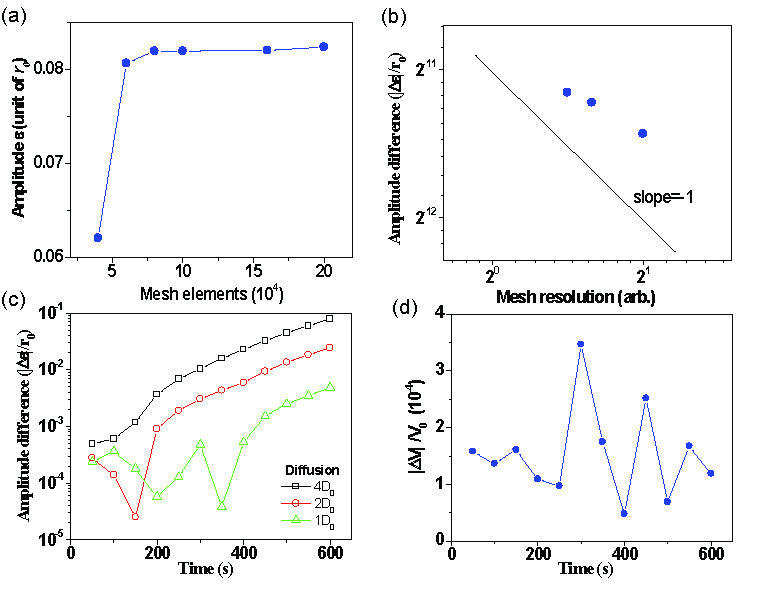}

\caption{\label{fig:convergence}Numerical convergence with respect to various
approximations in the simulation. (a) Numerical convergence shown
by instability amplitude at time t=300 seocnds vs number of mesh elements;
(b) established errors ($|\Delta\varepsilon|$ corresponding to higher
resolution) versus resolution (solid line=expected asymptotic rate);
(c) the artificial diffusion term for numerical stability, and (d)
the numerical errors represented by deviations from volume conservation. }

\end{figure}

\textcolor{black}{In order to establish the accuracy of the simulation,
we first investigated the convergence with respect to the various
approximations: the time integration tolerance ($\rightarrow0$),
the mesh resolution ($\rightarrow\infty$), and an artificial diffusion
term $\mathbf{\mathrm{D}}$ ($\rightarrow0$). }The time integration
is performed by a fifth-order backward differentiation formula, characterized
by specified relative and absolute error tolerances ($\mathrm{err_{rel}}=10^{-4}$
and $\mathrm{err_{abs}=10^{-5}}$) with acceptable accuracy ($\approx10^{-3}$).
Fig. \ref{fig:convergence}(a) shows numerical convergence of the
instabiltiy amplitude $\min_{z}\varepsilon_{1}(z,t)$ at time t=300
seocnds as a fucntion of the number of triangular mesh elements. From
last four data points in (a), the convergence of $|\Delta\varepsilon|/r_{0}$
with respect to the mesh resolution, compared with a very fine mesh
of $20\cdot10^{4}$ mesh elements, is presented in Fig. \ref{fig:convergence}(b).
Because of the PDE's discontinuous coefficient, we expect 1st-order
convergence: that is error $\sim$ 1/(mesh resolution), corresponding
to slope $-1$ (solid line) in Fig. \ref{fig:convergence}(b): the
data points appear to be approaching this slope, but we are not able
to measure errors at high enough resolution to verify the convergence
rate conclusively. However, we found $10\cdot10^{4}$ mesh elements,
corresponding to a typical element diameter $\approx0.01r_{0}$ ,
yield good accuracy ($\approx10^{-3}$ error). The physical results
are recovered in the limit $D\rightarrow0$, and so we must establish
the convergence of the results as $D$ is reduced and identify a $D$
small enough to yield accurate results while still maintaining stability.
\textcolor{black}{Fig. \ref{fig:convergence}(c) demonstrates this
convergence in the simulation by calculating the amplitude difference
$|\Delta\varepsilon|/r_{0}$ compared with $D=0.5D_{0}$ ($D_{0}=10^{-14}\mathrm{m^{2}/s}$
in Appendix 1), and 3 digits of accuracy are attained by $D=D_{0}$.
Accordingly, the numerical parameters throughout the following simulations
are $err_{rel}=10^{-4},err_{abs}=10^{-5},D=1.0D_{0}$, and $10\cdot10^{4}$
mesh elements. As an additional check, the volume of the incompressible
fluid shell between the two interfaces should be conserved. As shown
in Fig. \ref{fig:convergence}(d), the numerical errors in the simulation
leads to only a small volume deviation, ($|V(t)-V(0)|/V(0)=|\Delta V(t)|/V(0)$),
of less than $0.4\%$. We conclude that the finite-element level-set
approach is numerically reliable for the simulation of capillary instabilities. }

\subsection{Instability evolution}

\begin{figure}
\includegraphics[width=1\columnwidth]{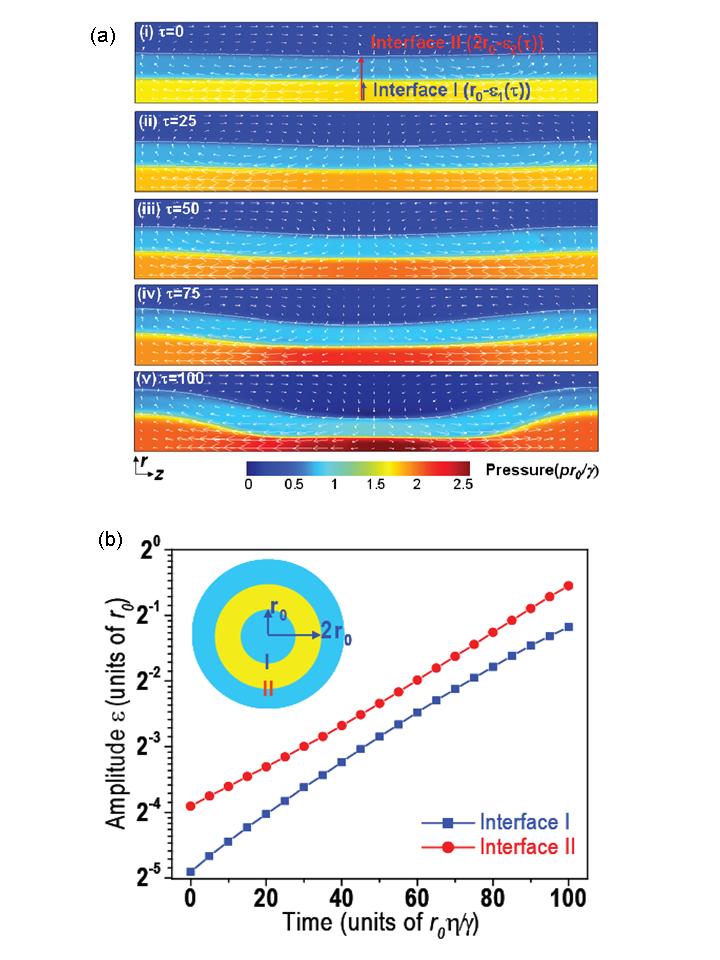}

\caption{\label{fig:evolution}(a) Snapshot of the flow field and interfaces
during instability evolution. Color scale for pressure, arrows for
fluid velocity. (b) The instability amplitude grows exponentially
with time at interfaces $\mathrm{I}$ and $\mathrm{II}$. }

\end{figure}

\textcolor{black}{The evolution of a capillary instability can be
obtained from direct numerical simulation.} Fig. \ref{fig:evolution}(a)
(i)--(v) presents snapshots of the $ $\textcolor{black}{{} flow field
and interface.} (i) Initially, the pressure of the inner fluid is
higher than that of the outer fluid due to Laplace pressure ($p=\gamma\kappa$)
originating from azimuthal curvature of the cylindrical geometry at
interfaces $\mathrm{I}$ and $\mathrm{II}$. (ii)--(iv) The interfacial
perturbations generate an axial pressure gradient $\Delta p$, and
hence a fluid flow occurs that moves from a smaller-radius to a larger-radius
region for the inner fluid. Gradually the amplitude of the perturbation
is amplified. (v) The\textcolor{black}{\emph{\noun{ }}}\textcolor{black}{\emph{shrunk}}\textcolor{blue}{\emph{
}}smaller-radius and \emph{expanded} larger-radius regions of inner
fluid further enhance the axial pressure gradient $\Delta p$, resulting
in a larger amplitude of the perturbation. As a result, the small
perturbation is exponentially amplified by the axial pressure gradient. 

Fig. \ref{fig:evolution}(b) shows the instability amplitude \textcolor{black}{$\min_{z}\varepsilon(z,t)$}
at interfaces $\mathrm{I}$ and $\mathrm{II}$ growing with time exponentially
on a semilog scale in the plot.\textcolor{blue}{{} }\textcolor{black}{$ $
The instability time scale in the simulation is about $372\pm3$ $\mathrm{sec}$
by fitting the curve to an exponential $\varepsilon\sim e^{t/\tau}$.
Linear theory predicts time scale of $\tau\approx334$ $\mathrm{sec}$
\cite{Stone1996} (see Appendix 2). Thus, simulation is consistent
with the linear theory, in terms of the instability time scale.}

\subsection{\textcolor{black}{Beyond the linear theory }}

\begin{figure}
\includegraphics[width=1\columnwidth]{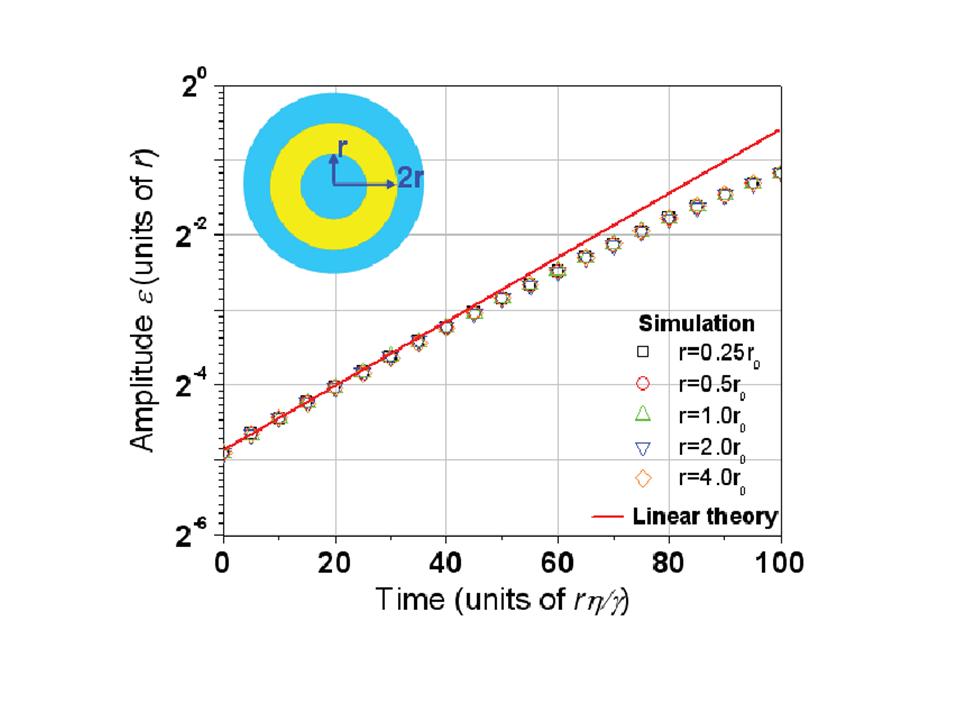}

\caption{\label{fig:scaling}Growth of perturbation amplitude as a function
of time. At small amplitudes below $10\%$ \textit{$r$}, the perturbation
grows exponentially as expected from linear theory. At large amplitude
above $10\%$ $r$, significant deviations from linear theory occur.
Scale invariance, with various values of r superimposed, is expected
in the Stokes regime with low Re number. }

\end{figure}

\textcolor{black}{More quantitatively, for small perturbations or
short times, exponential growth of the perturbation amplitude is expected
from the linear theory. However, at later times, when the perturbation
has grown sufficiently, one expects the linear theory to break down.
We observe significant deviations from the linear theory if the amplitude
is above $10\%$ of the cylinder radius $r$ (corresponding to time
$40r\eta/\gamma$), as shown in Fig. \ref{fig:scaling} (taking interface
$\mathrm{I}$ as an example). For example, at perturbation amplitudes
of around $25\%$ of $r$ (time $70r\eta/\gamma$), the deviations
of the simulation data from the linear theory is almost $20\%$ . }

\textcolor{black}{The length scaling of the instability is studied
as well. After rescaling time and distance by the same factor (so
that the initial radius is scaled by $0.25,0.5,1,2$, and $4$), all
the data collapse onto a single master curve, as shown in Fig. \ref{fig:scaling}.
This length scaling originates from the low $\mathrm{Re}$ regime,
which implies that the NS equations are approximately linear with
respect to $\vec{u}$; this gives rise to the well known scale invariance
in the Stokes regime \cite{Batchelor00}. }

\subsection{Unequal viscosities\label{sub:Unequal-viscosities} }

\begin{figure}
\includegraphics[width=1\columnwidth]{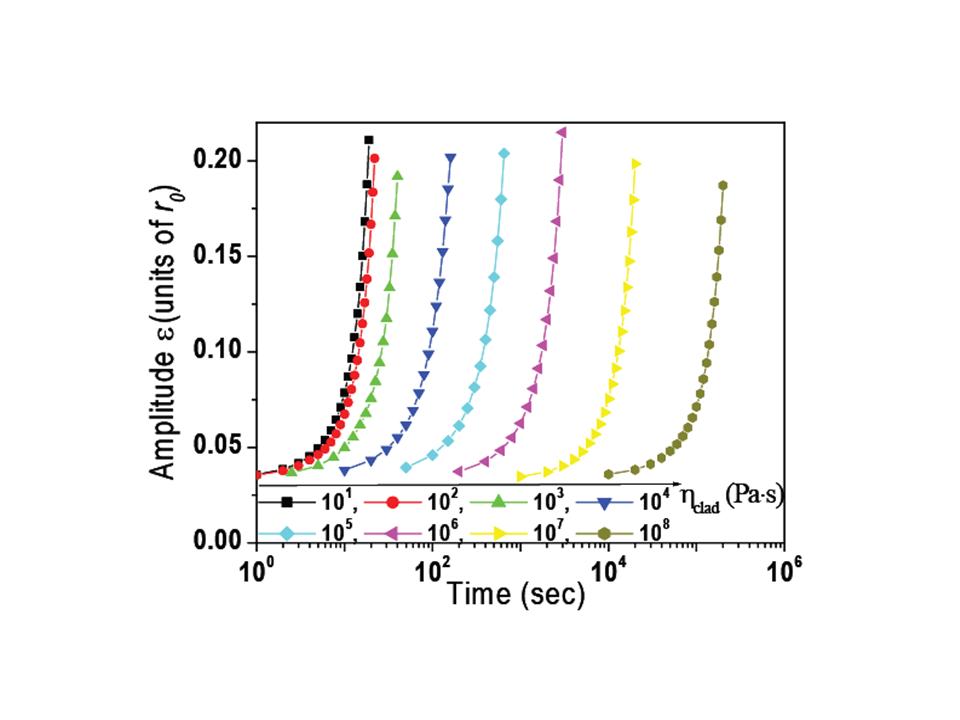}

\caption{\label{fig:viscositycontrast}Time-dependent perturbation amplitude
curves for various cladding viscosities ($\eta_{\mathrm{clad}}$),
but with a fixed shell viscosity ($\eta_{\mathrm{shell}}=10^{5}\mathrm{Pa\cdot s}$). }

\end{figure}

\begin{figure}
\includegraphics[width=0.9\columnwidth]{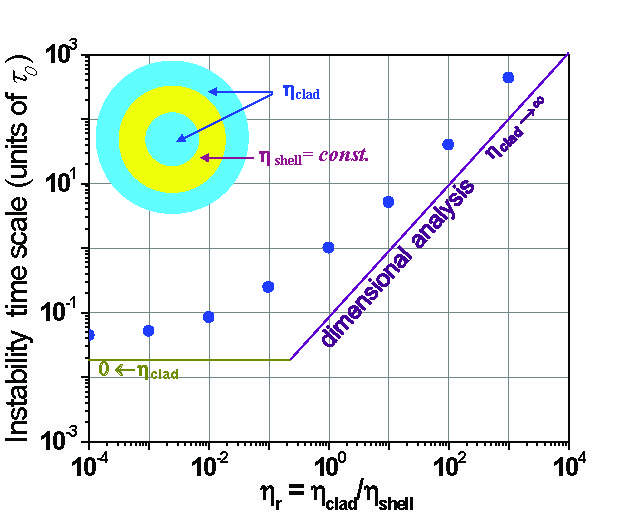}

\caption{\label{fig:viscositycladding}An instability time scale ($\tau$)
for unequal viscosity with a fixed shell viscosity ($\eta_{\mathrm{shell}}$).
In the limit of $\eta_{\mathrm{clad}}\rightarrow0$, $\tau$ is determined
by $\eta_{\mathrm{shell}}$ and approaches to a constant. In the opposite
limit of $\eta_{\mathrm{clad}}\rightarrow\infty$, $\tau$ should
be determined by $\eta_{\mathrm{clad}}$ and is linearly proportional
to $\eta_{\mathrm{clad}}$. \textcolor{black}{Between these two limits
of }$\eta_{\mathrm{clad}}$\textcolor{black}{, }$\tau$\textcolor{black}{{}
$ $ smoothly interpolates between the corresponding time scales. }}

\end{figure}

We now investigate the dependence of instability timescale on cladding
viscosity ($\eta_{\mathrm{clad}}$) with a fixed shell viscosity ($\eta_{\mathrm{shell}}$),
in order to help us to identify suitable cladding materials for fiber
fabrication. As viscosity slows down fluid motion, the low-viscosity
cladding has a faster instability growth rate and a shorter time scale,
while the high-viscosity cladding has a slower instability growth
rate and a longer instability time scale. Fig. \ref{fig:viscositycontrast}
shows the time-dependent perturbation amplitude curves for various
viscosity contrast $\eta_{\mathrm{clad}}/\eta_{\mathrm{shell}}$ by
changing the cladding viscosity ($\eta_{\mathrm{shell}}=10^{5}$ $\mathrm{Pa\cdot s}$).
Instability time scale for the each given viscosity contrast is obtained
by exponentially fitting the curves in Fig. \ref{fig:viscositycontrast}.
Instability time scale ($\tau$) as a function of viscosity contrast
is presented in Fig. \ref{fig:viscositycladding}\textcolor{black}{. }

The existing linear theory has only been solved in case of equal viscosity,
and predicts that the instability time scale is proportional to the
viscosity $\tau\sim\eta$ \cite{Stone1996}. We obtain a more general
picture of the instability time scale for unequal viscosity by considering
two limits. In the limit of \textcolor{black}{negligible cladding
viscosity, $\eta_{\mathrm{clad}}\rightarrow0$, the instability time
scale should be determined by $\mathrm{\mathrm{\eta}}_{\mathrm{shell}}$,
and from dimensional analysis should be proportional to }$r\eta_{\mathrm{shell}}/\gamma$
,\textcolor{black}{{} assuming that the inner and outer radius are comparable
and so we take $r$ to be the average radius. In the opposite limit
of }$\eta_{\mathrm{clad}}\rightarrow\infty$ \textcolor{black}{, the
time scale should be determined by }$\eta_{\mathrm{clad}}$ \textcolor{black}{and
hence should be proportional to} $r\eta_{\mathrm{clad}}/\gamma$\textcolor{black}{.
In between these two limits, we expect the time scale to smoothly
interpolate between the }$r\eta_{\mathrm{shell}}/\gamma$\textcolor{black}{{}
$ $ and }$r\eta_{\mathrm{clad}}/\gamma$\textcolor{black}{{} $\mathrm{}$
scales. $ $ Precisely this behavior is observed in our numerical
calculations, as shown in Fig. \ref{fig:viscositycladding}. }

\section{\textcolor{black}{Estimate of radial instability timescale\label{sec:Radial-Stability-Map}}}

\textcolor{black}{In the previous section, we surveyed the capillary
instability of a concentric-cylindrical shell by numerical simulations.
We proceed to apply these results to investigate the instability time
scale dependent on various physical parameters including geometry
(e.g., radius and shell thickness) and materials properties (viscosity)
in a broad range, exploiting the accuracy of simple dimensional analysis
demonstrated in the previous section and providing guidance of our
fiber drawing. }

\begin{figure}
\includegraphics[width=1\columnwidth]{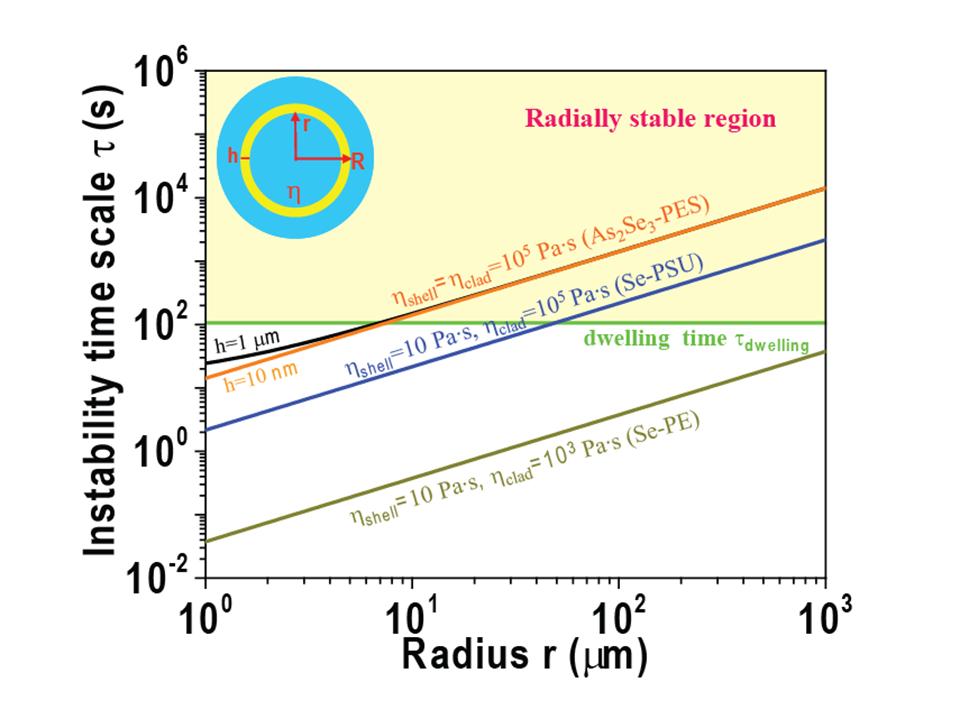}

\caption{\label{fig:Stabilitymap}Radial stability map. Linear theory calculations
of the instability time scale ($\tau$), which is dependent on the
radius, thickness, and viscosity. Inset shows cross-sectional geometry
of cylindrical shell. In our experiments, the dwelling time of thermal
drawing is around $\tau_{\mathrm{dwelling}}\approx100$ $\mathrm{sec}$,
and the fiber radius $r\approx500$ $\mathrm{\mathrm{\mu m}}$. Radially
stable region is shaded yellow for $\tau>\tau_{\mathrm{dwelling}}$,
while the unstable region corresponds to $\tau<\tau_{\mathrm{dwelling}}$. }

\end{figure}

\textcolor{black}{The calculated }instability time scale ($\tau$)
for different values of the radius ($r$) and the viscosity ($\eta$)
is displayed in Fig. \ref{fig:Stabilitymap}. \textcolor{black}{The
cross-s}ectional geometry in the calculation is shown in the inset:
interface \textit{\textcolor{black}{$\mathrm{I}$$ $}}\textcolor{blue}{{}
}$ $ is located at radius $r$, the cylindrical-shell thickness is
$h$, and interface \textit{\textcolor{black}{$\mathrm{II}$}} is
at radius $R=r+h$. The interfacial tension in the calculations was
set to \textcolor{black}{$\gamma=0.1\mbox{}\mathrm{N/m}$}, which
was the measured interfacial tension \textcolor{black}{between thermoplastic
polymer and chalcogenide glass used in our microstructured fibers
\cite{Hart2004}}. 

Two cases are considered: one is equal viscosity $(\eta_{\mathrm{shell}}=\eta_{\mathrm{clad}})$,
and the other is unequal viscosity $(\eta_{\mathrm{shell}}\neq\eta_{\mathrm{clad}})$.
In the case of $\eta_{\mathrm{shell}}=\eta_{\mathrm{clad}}$, the
instability time scale is calculated exactly from Stone and Brenner's
linear theory \cite{Stone1996}, 

\begin{equation}
\tau=\frac{\eta r}{\gamma\max_{\lambda}\Psi(\lambda,R/r)},\label{eq:equvistimescale}\end{equation}
where the fastest growth factor $\mathrm{\max_{\lambda}}\Psi(\lambda,R/r)$
was found by searching numerically within a wide range of wavelengths
$\lambda$ for a certain value of $R/r$ (Eq. \ref{eq:Growthrate}
in Appendix 2). Fig. \ref{fig:Stabilitymap} plots this time scale
versus radius for $\mathrm{\eta=10^{5}\mbox{}\mbox{}Pa\cdot s}$ corresponding
to As$_{2}$Se$_{3}$--PES, compared to the dwelling time \textcolor{black}{$\tau_{\mathrm{dwelling}}\approx100$
$\mathrm{sec}$ which is defined by the time of materials in viscous
state before exiting hot furnace to be frozen in fiber during thermal
drawing \cite{Hartthesis}}.

In the other case of $\eta_{\mathrm{shell}}\neq\eta_{\mathrm{clad}}$
(in the $\eta_{\mathrm{clad}}/\eta_{\mathrm{shell}}\gg1$ regime as
discussed in Section \ref{sub:Unequal-viscosities}), the instability
time scale can be roughly estimated from dimensional analysis. Although
dimensionless analysis does not give the constant factor, for specificity,
we choose the constant coefficient from the Tomotika model \cite{Tomotika1935},

\begin{equation}
\tau\approx\frac{2r\eta_{clad}}{\gamma\max_{\lambda}\left[(1-x^{2})\Phi(x,\eta_{clad}/\eta_{shell})\right]},\label{eq:unequtimescale}\end{equation}
where the fastest growth factor $\mathrm{\max_{\lambda}[}(1-x^{2})\Phi(x,\eta_{clad}/\eta_{shell})]$
was found numerically by searching a wide range of wavelengths ($x=2\pi r/\lambda$)
{[}$\mathrm{\Phi(x,\eta_{clad}/\eta_{shell})}$ is a complicated implicit
function of wavelength and viscosity contrast given in Ref. \cite{Tomotika1935}{]}.
Fig. \ref{fig:Stabilitymap} plots the time scale for $\eta_{\mathrm{shell}}=10$
$\mathrm{Pa\cdot s}$, $\mathrm{\mathrm{\eta}_{clad}=10^{5}}$ $\mathrm{Pa\cdot s}$,
corresponding to Se--PSU showing that the observed stability of shells
of radius $\mathrm{\approx250\mbox{}\mu m}$ is consistent with the
radial stability criterion ($\mathrm{\tau>\tau_{dwelling}}$). On
the other hand, if $\mathrm{\eta_{clad}}$ is reduced to $\mathrm{10^{3}}$$ $
$\mathrm{Pa\cdot s}$ with the same shell materials, corresponding
to Se--PE, we predict that radial fluctuations alone will render the
shell unstable for any radius $r\leq1$ $\mbox{}$$\mathrm{mm}$.

\section{Applications in Microstructured Fibers \label{sec:Applications-in-Microstructured}}

\textcolor{black}{In this section, we consider in more detail the
application of these analyses to understand observed experimental
results, and in particular the observed stability (or instability)
of thin shells and filaments. We examine whether our stability analysis
can provide guidance in materials selection and in the the understanding
of attainable feature sizes. Because we only considered radial fluctuations,
our analysis provides a necessary but not sufficient criterion for
stability. Therefore, the relevant questions are whether the criterion
is }\textcolor{black}{\emph{consistent}}\textcolor{black}{{} $ $ with
observed stable structures, whether it is sufficient to explain the
observed azimuthal breakup, and what materials combinations are }\textcolor{black}{\emph{excluded}}\textcolor{black}{.
Below, in Sec \ref{sub:Comparision-with-observations} we consider
the application of radial stability analysis to the observed stability
or instability of cylindrical shells. In Sec \ref{sub:Materials-selection}
we look at the impact on materials selections. Finally, in Sec \ref{sub:Discussion-of-continuous},
we show that the observed stability of the resulting nanoscale filaments
is consistent with the Tomotika model. }

\subsection{Comparison with observations for cylindrical shells\label{sub:Comparision-with-observations}}

The radial stability map of Fig. \ref{fig:Stabilitymap} is consistent
with the experimental observations (the default cylindrical shell
radius $\approx250$$ $$\mathrm{\mu m}$). First, the map predicts
that feature sizes down to submicrometers and hundreds of nanometers
are consistent with radial stability for the equal-viscosity materials
combination of $\eta_{\mathrm{clad}}=\eta_{\mathrm{shell}}=10^{5}$
$\mathrm{Pa\cdot s}$, which corresponds to As$_{2}$Se$_{3}$--PES
or As$_{2}$S$_{3}$--PEI. For As$_{2}$Se$_{3}$--PES, Fig.\ref{fig:SEM}
(c) shows that a shell thickness of As$_{2}$S$_{3}$ of $1$ $\mathrm{\mathrm{\mu m}}$
is obtained; in other work, layers of As$_{2}$S$_{3}$ down to $15$
$\mathrm{nm}$ have been achieved as well \cite{Deng2008}. For As$_{2}$S$_{3}$--PEI,
Fig.\ref{fig:SEM} (d) demonstrates a thickness of $\mathrm{As_{2}S_{3}}$
down to 32 $\mathrm{nm}$. Second, the map is consistent with thicknesses
down to submicrometers for unequal-viscosity materials with $\eta_{\mathrm{clad}}=10^{5}$
$\mathrm{Pa\cdot s}$, $\eta_{\mathrm{shell}}=10$ $\mathrm{Pa\cdot s}$,
which corresponds to Se--PSU. Se layers with thickness on the order
of $1$ $\mathrm{\mathrm{\mu m}}$ have been demonstrated in Se--PSU
fiber \cite{Deng2008}. 

\textcolor{black}{The radial stability map, nevertheless, is not sufficient
to explain the azimuthal instability at further reduced thicknesses
down to tens of nanometers. }The stability map of Fig. \ref{fig:Stabilitymap}
predicts that a $\mathrm{Se}$ layer in a Se--PSU combination$ $
should be \emph{radially} stable down to tens of nanometers\textcolor{black}{.
However, we found in experiments {[}Fig. \ref{fig:SEM}(e, f){]} that
a Se shell with thickness $<100$ nm breaks up into continuous filament
arrays }\cite{Deng2008,Deng2009}\textcolor{black}{, which means that
the mechanism for this breakup is distinct from that of purely radial
fluctuations. }For another As$_{2}$Se$_{3}$--PES materials combination,
this filamentation of As$_{2}$Se$_{3}$ film was also observed as
the thickness is reduced down to $10$ $\mathrm{nm}$ \cite{Deng2008}.\textcolor{black}{{}
$ $ Future work will elucidate this filamentation mechanism by performing
$\mathrm{3D}$ numerical simulation to explore the azimuthal fluctuations. }

\subsection{Materials selection\label{sub:Materials-selection}}

\begin{figure}
\includegraphics[width=1\columnwidth]{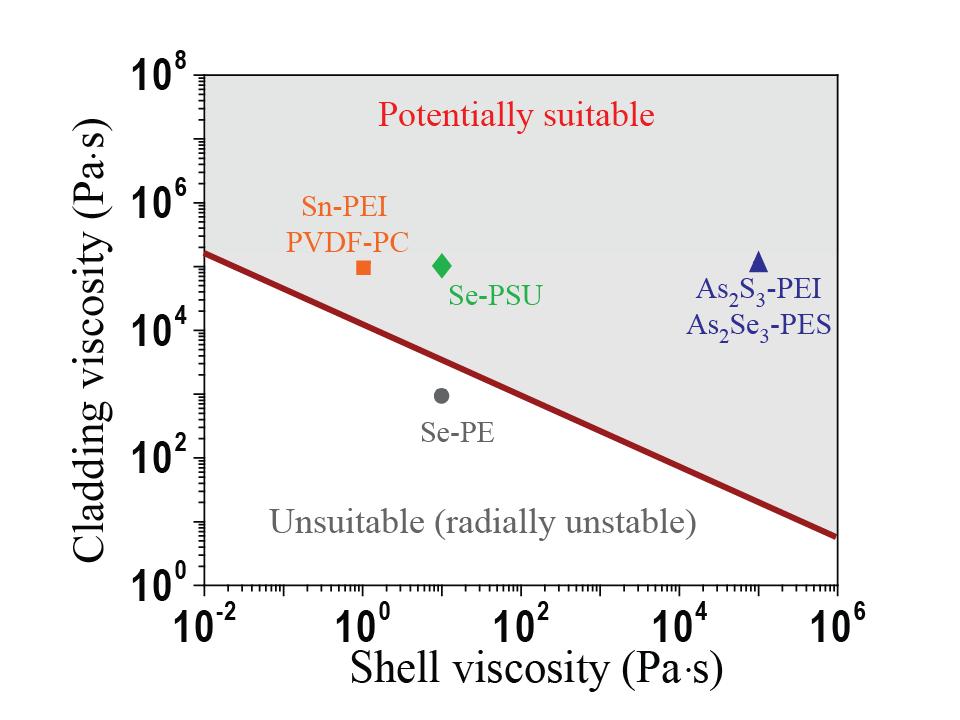}

\caption{\label{fig:Identified-viscous-materials} Calculated shell--cladding
viscous materials selection map during thermal drawing ($\tau_{\mathrm{dwelling}}=100$
sec). A red line for instability time for dwelling time $\tau=\tau_{\mathrm{dwelling}}$.
The shaded region above the red line indicates potentially suitable
viscous materials combination ($\tau>\tau_{\mathrm{dwelling}}$),
those in which radial instabilities alone do not cause breakup. The
region below the red line indicates radially unstable materials combinations
($\tau<\tau_{\mathrm{dwelling}}$), such as $\mathrm{Se-PE}$ materials
combination, which are unstable for thermal drawing. }

\end{figure}

Given the viscosities and surface tension of a particular material
pair, we can use Fig. \ref{fig:Stabilitymap} to help determine whether
that pair is suitable for drawing: if it is radially unstable, then
it is almost certainly unsuitable (unless the process is altered to
somehow compensate), whereas if it is radially stable then the pair
is at least potentially suitable (if there are no other instabilities).
\textcolor{black}{Viscosities of materials used in fiber drawing are
obtained as follows: the viscosities of semiconductor glasses (Se,
As$_{2}$Se$_{3}$, As$_{2}$S$_{3}$) are calculated from }an empirical
Arrhenius formula\textcolor{black}{{} $ $ at the associated temperature
during thermal drawing (more details in Appendix 3); }several thermoplastic
polymers (PSU, PES, and PEI) have similar viscosities $\eta_{\mathrm{polymer}}\approx10^{5}$
$\mathrm{Pa\cdot s}$ during fiber drawing \cite{Hartthesis}; and
the viscosity\textcolor{black}{{} of the polymer PE is $10^{3}$ $\mathrm{Pa\cdot s}$
at temperature $\mathrm{T=250^{\circ}C}$ \cite{Dobrescu83}. }The
polymer--glass surface tension is typically $\gamma=0.1$$ $ N/m
\cite{Hart2004} for all of these materials. Assuming a cylindrical
shell of radius $\approx250$ $\mathrm{\mu m}$ and a dwelling time
of thermal drawing  $\approx100$ $\mathrm{sec}$\textcolor{black}{,
we can classify each materials combination by whether it falls in
the $\tau>\tau_{\mathrm{dwelling}}$ yellow region (radially stable)
of Fig. \ref{fig:Stabilitymap} or in the $\tau<\tau_{\mathrm{dwelling}}$
white region (radially unstable). }

These materials combinations are presented in Fig. \ref{fig:Identified-viscous-materials}.
The boundary line in red, which indicates the viscous combinations
that satisfy \textcolor{black}{$\tau=\tau_{\mathrm{dwelling}}$, divides
the map into two areas.} The shaded area above the boundary line is
the region of potentially suitable materials combinations for fiber
drawing (As$_{2}$Se$_{3}$--PES, As$_{2}$S$_{3}$--PEI, Se--PSU)
\cite{Abouraddy2007,Deng2008}; while the materials combinations below
the boundary line are unsuitable due to radial instability. Here,
the only materials combination which seems to be definitely excluded
in this regime due to radial instability is Se--PE. The polymer--polymer
materials combination of PVDF--PC (polyvinylidene fluoride, PVDF,
a piezoelectric polymer; polycarbonate, PC) is potentially suitable
for thermal drawing, and this possibility has been confirmed by recent
experiments \cite{Egusa}. \textcolor{black}{Moreover, since a high
viscosity cladding improves stability (increase $\tau$), we predict
that a wider variety of shell materials with low viscosity may possibly
be employed in microstructured fibers, such as the metals $\mathrm{Sn}$
and $\mathrm{\mathrm{I}n}$ \cite{Abouraddy2007,Culpin1957}. These
various available classes of metals, polymers and semiconductors expand
the potential functionalities of devices in microstructured fibers. }

\subsection{Stability of continuous filaments down to submicrometer/nanometer
scale\label{sub:Discussion-of-continuous}}

As shown in Fig. \ref{fig:SEM}(e--f), in some cases the initial shell
breaks up (azimuthally) into long cylindrical filaments \cite{Deng2008,Deng2009}.
These filaments themselves should be subject to the classic capillary
instability and in principle should eventually break up into droplets.
In our fiber-drawing experiments, however, no further instability
is observed: the filaments are observed to remain continuous and unbroken
over at least cm length scales with diameters reaching submicrometer
scales \cite{Deng2008,Deng2009}. Since a cylindrical filament of
one fluid surrounded by another should be described exactly by the
Tomotika model, it is important to know whether the observed stability
is consistent with this model, or otherwise requires some additional
physical mechanism (such as visco-elasticity) to explain. In this
section, we find that the timescale of instability predicted by the
Tomotika model exceeds the dwelling time of fiber drawing, making
it unsurprising that filament instability is not observed. In fact,
we show that the instability time scale exceeds the dwelling time
even under unrealistically conservative assumptions: that the filaments
appear immediately in the fiber drawing, that the maximum instability
growth rate is cumulative even though the lengthscale achieving maximum
growth changes with the filament radius during drawing, and that the
polymer viscosity is always at its minimum value (corresponding to
the highest temperature point). 

An instability with time scale $\tau$ corresponds to exponential
growth of a fluctuation amplitude $\epsilon$ according to $\frac{d\epsilon}{dt}=\epsilon/\tau$.
If $\tau$ is time varying, then the total amplitude growth is $\mathrm{exp}[\int dt/\tau(t)]$.
Converting to $dz=v(z)dt$ for a position-dependent axial flow velocity
$v(z)$ during thermal drawing, we therefore obtain a total exponential
growth factor: 

\begin{equation}
\Gamma=\int_{0}^{L}\frac{dz}{v(z)\tau(z)},\label{eq:dimensionlesscritia}\end{equation}
where $z\in[0,L]$ is axial position in the neck-down region with
length $(\mathrm{L=6\mbox{ \ensuremath{}}}\mathrm{cm})$.  $\Gamma\gg1$
corresponds to breakup, while $\Gamma\ll1$ corresponds to stability. 

In order to provide a conservative estimation of filament stability,
the capillary instability time is calculated from the fastest growth
factor at each axial location (this is a very conservative estimate),
and the polymer viscosity is set to be the minimum value (at the highest
temperature) during thermal drawing. The capillary instability time
scale is calculated based on the Tomotika model as follows \cite{Tomotika1935},

\begin{equation}
\tau(z)=\frac{2r\eta_{\mathrm{polymer}}(z,T)}{\gamma\left\{ \max_{\lambda}\left(1-x^{2}\right)\Phi\left[x,\eta_{\mathrm{polymer}}(z,t)/\eta_{\mathrm{clad}}(z,t)\right]\right\} }.\label{eq:Tomotikacal}\end{equation}
The complex shape of neck-down profile is fitted from experiment can
be approximately described by following formula,

\begin{equation}
\frac{R(z)}{R(0)}=(1+k\frac{z}{L})^{-1/p},\quad k=\left[\frac{R(0)}{R(L)}\right]^{p}-1,\quad p=2.\label{eq:neckdownprofile}\end{equation}
Due to the incompressibility of the viscous fluid, the velocity of
flow scales inversely with area: 

\begin{equation}
\frac{v(z)}{v(0)}=\frac{R^{2}(0)}{R^{2}(z)},\label{eq:velocity}\end{equation}
where $\mathrm{v(0)=4\times10^{-3}\mbox{}mm/sec}$$ $ is the preform
velocity. Again by incompressibility, the filament radius $(\mathrm{r})$
should scale as the fiber radius $(R)$:

\begin{equation}
\frac{r(z)}{r(0)}=\frac{R(z)}{R(0)}.\label{eq:filamentradius}\end{equation}
The temperature distribution during thermal drawing, fit from experiment,
is found to be approximately parabolic,

\begin{equation}
T=T_{\mathrm{max}}-\left(T_{\mathrm{max}}-T_{\mathrm{min}}\right)\left(2\frac{z}{L}-1\right)^{2}.\label{eq:temperature}\end{equation}

\begin{figure}
\includegraphics[width=1\columnwidth]{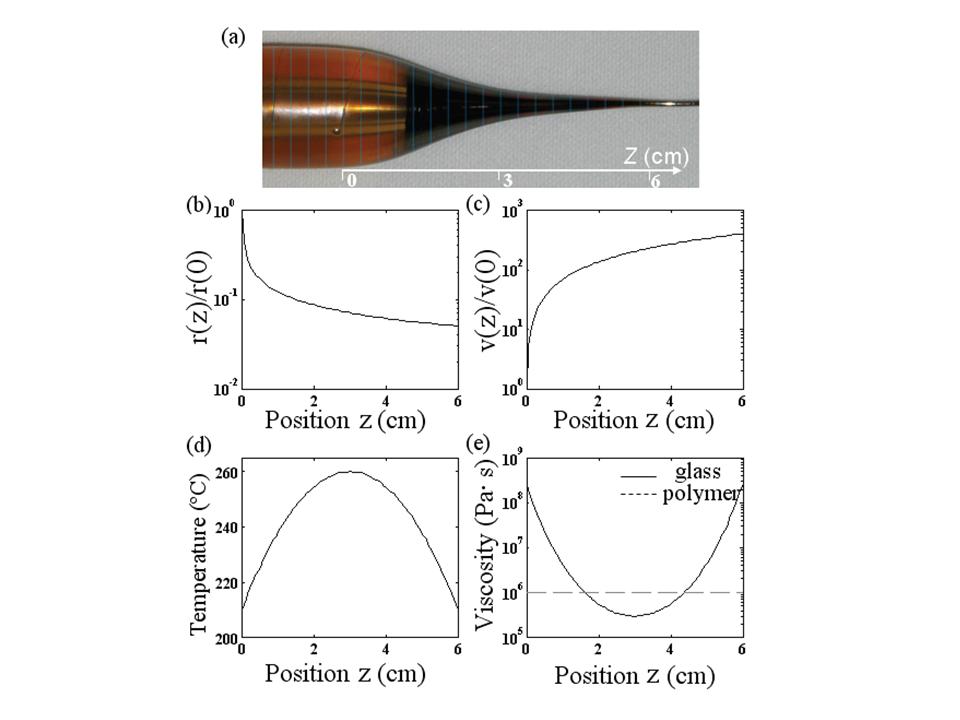}

\caption{\label{fig:Calculated-relevant-parameters}Relevant parameters in
the neck-down region during thermal drawing. (a) Photograph of neck-down
region from preform to fiber, (b)--(e) for the calculated radius,
velocity, temperature and viscosity. }

\end{figure}

In calculations, parameters for the typical $\mathrm{As_{2}Se_{3}}$--PES
fiber drawing are $\mathrm{R(0)=1}$~cm,$ $ $\mathrm{s=20}$, $\mathrm{L=6}$~cm,
$\mathrm{p=2}$, $ $ $\mathrm{T_{max}=260}\,{}^{\circ}$C, $\mathrm{T_{min}=210}\,{}^{\circ}$C,
$ $ $\mathrm{r(L)=200}$~nm, $\mathrm{\eta_{polymer}=10^{6}}$~Pa$\cdot$s.
Fig. \ref{fig:Calculated-relevant-parameters} (b)--(d) presents the
corresponding position--dependent variables including radius, velocity,
temperature and viscosity. Finally, we obtain 

\begin{equation}
\Gamma=0.90.\label{eq:dimensionlesssnumber}\end{equation}
This satisfies $\Gamma<1$, but only barely---if this were an accurate
estimate of the growth factor, instability might still be observed.
However, the assumptions we made above were so conservative that the
true growth factor must be much less than this, indicating the instability
should not be observable during the dwelling time of fiber drawing.
So, the observed filaments are consistent with the Tomotika model,
although of course we cannot yet exclude the possibility that there
are also additional effects (\textit{e.g.,} elasticity) that further
enhance stability. 

\subsection{Favorability of azimuthal versus axial instability \label{sub:azimuthal-preference}}

In experiments, we observed that thin films preferentially break up
along the azimuthal direction rather than the axial direction. The
discussion of the previous Section~\ref{sub:Discussion-of-continuous}
suggests a simple geometrical explanation for such a preference, regardless
of the details of the breakup mechanism. The key point is that any
instability will have some characteristic wavelength $\lambda$ of
maximum growth rate for small perturbations, and this $\lambda$ must
be proportional to the characteristic feature size of the system,
in this case the film thickness $d$. As the fiber is drawn, however,
the thickness $d$ and hence $\lambda$ decreases. Now, we consider
what happens to an unstable perturbation that begins to grow at some
wavelength $\lambda_{0}$ when the thickness is $d_{0}$. If this
is a perturbation along the \emph{axial} direction, then the fiber-draw
process will \emph{stretch} this perturbation to a \emph{longer} wavelength,
that will no longer correspond to the maximum-growth $\lambda$ (which
is shrinking), and hence the growth will be damped. That is, the axial
stretching competes with the layer shrinking, and will tend to suppress
\emph{any} axial breakup process. In contrast, if $\lambda_{0}$ is
an \emph{azimuthal} perturbation, the draw-down process will \emph{shrink}
$\lambda_{0}$ along with the fiber cross-section at exactly the same
rate that $d$ and $\lambda$ shrink. Therefore, azimuthal instabilities
are \emph{not} suppressed by the draw process. This simple geometrical
argument immediately predicts that the first observed instabilities
will be azimuthal (although axial instabilities may still occur if
the draw is sufficiently slow).

\section{Concluding remarks }

In this paper, motivated by recent development in microstructured
optical fibers, we have explored capillary instability due to radial
fluctuations in a new geometry of concentric cylindrical shell by
2D numerical simulation, and applied its theoretical guidance to feature
size and materials selections in the microstructured fibers during
thermal drawing processes. Our results suggest several directions
for future work. First, it would be desirable to extend the analytical
theory of capillary instability in shells, which is currently available
for equal viscosity only, to the more general case of unequal viscosities
--- we have developed a very general theory in an appearing paper
\cite{Liang2010}. Second, we plan to extend our computation simulations
to include 3D azimuthal fluctuations together with radial fluctuations;
as argued in Section~\ref{sub:azimuthal-preference}, we anticipate
a general geometrical preference for azimuthal breakup over axial
breakup once the draw process is included. Third, there are many additional
possible experiments that would be interesting to explore different
aspects of these phenomena in more detail, such as employing different
geometries (\textit{e.g.}, non-cylindrical), temperature-time profiles,
or materials (\textit{e.g.}, Sn--PEI or Se--PE). Finally, by drawing
more slowly so that axial breakup occurs, we expect that experiments
should be able to obtain more diverse structures (\textit{e.g.} axial
breakup into rings or complete breakup into droplets) that we hope
to observe in the future.

\section*{Acknowledgments}

This work was supported by the Center for Materials Science and Engineering
at MIT through the MRSEC Program of the National Science Foundation
under award DMR-0819762, and by the U.S. Army through the Institute
for Soldier Nanotechnologies under contract W911NF-07-D-0004 with
the U.S. Army Research Office.

\section*{Appendix }

\section*{1. Direct numerical simulation of concentric cylindrical shells }

In the simulation, the level set function is defined by a smoothed
step function reinitialized at each time step \cite{Olsson2005}, 

\begin{equation}
\phi(z,t)=\begin{cases}
0 & r<r_{1}(z,t)-\Lambda,\enskip r>r_{2}(z,t)+\Lambda\\
1 & r_{1}(z,t)+\Lambda<r<r_{2}(z,t)-\Lambda\\
0.5 & \mbox{otherwise}\end{cases},\label{eq:InitialLS}\end{equation}
where $r_{1}(z,t)$ and $r_{2}(z,t)$ are the radius of interfaces
I and II of the coaxial cylinder, and $\Lambda$ is the half thickness
of the discretized interface. The level set $\phi=1$, 0, and $0.5$
corresponds to the region of concentric cylindrical shell, of surrounding
fluid, and interface, respectively. Contour $\Pi=\{X|\phi(X,t)=0.5\}$
tracks the interface $\mathrm{I}$ and $\mathrm{II}$. 

A smoothed delta function is defined to project surface tension at
interface, 

\begin{eqnarray*}
\delta & = & 4\phi(1-\phi)=\begin{cases}
1 & |r-r_{1,2}|<\Lambda\\
0 & \mbox{otherwise}\end{cases}.\end{eqnarray*}
A smoothed step function is introduced to create a smooth transition
of the level-set function $\phi$ from $0$ to $1$ across the interface, 

\begin{equation}
\phi=\left[1+e^{(r-r_{1})/\Lambda}\right]^{-1}-\left[1+e^{(r-r_{2})/\Lambda}\right]^{-1}.\label{eq:Stepfunction}\end{equation}

The boundary condition for the level set equation at the edge of the
computational cell is 

\begin{equation}
\vec{n}\cdot(-c\nabla\phi+\phi\vec{u})=0,\label{eq:LSboundary}\end{equation}
where $\vec{n}$ and $\vec{t}$ are the normal and tangential vectors
at the boundary. In addition, boundary conditions for the NS equations
are 

\begin{equation}
\begin{cases}
\vec{n}\cdot\vec{u} & =0,\\
\vec{t}\cdot\eta\left[\nabla\vec{u}+(\nabla\vec{u})^{T}\right]\vec{n} & =0.\end{cases}\label{eq:NSboundary}\end{equation}

In the simulation, time-stepping accuracy is controlled by an absolute
and relative error tolerance ($\mathrm{err_{abs}}$ and $\mathrm{err_{rel}}$)
for each integration step. Let $U$ be the solution vector at a given
time step, $E$ be the solver\textquoteright{}s estimated local error
in $U$ during this time step, and $N$ $ $ be the number of degrees
of freedom in the simulation. Then a time step is accepted if the
following condition is satisfied, 

\begin{equation}
\left[\frac{1}{N}\underset{i}{\sum}\left(\frac{|E_{i}|}{err_{abs}+err_{rel}|U_{i}|}\right)^{2}\right]^{^{1/2}}<1.\label{eq:Numerical}\end{equation}

A triangular finite-element mesh is generated, and second-order quadratic
basis functions are used in the simulation. Parameters for Fig. \ref{fig:evolution}
in the simulation are $\rho=10^{3}$ $\mathrm{kg/m^{3}}$,$ $ $\eta=10^{5}$
$\mathrm{Pa\cdot s}$, $\gamma=0.6$ $\mathrm{N/m}$, $R=120$ $\mathrm{\mu m}$,
$\mathrm{err_{rel}}=10^{-4},\mathrm{\mbox{}err_{abs}}=10^{-5},$ $D=D_{0}=10^{-14}$
$\mathrm{m^{2}/s}$.

\section*{2. Linear theory of concentric  cylindrical shells with equal viscosities }

\begin{figure}
\includegraphics[width=1\columnwidth]{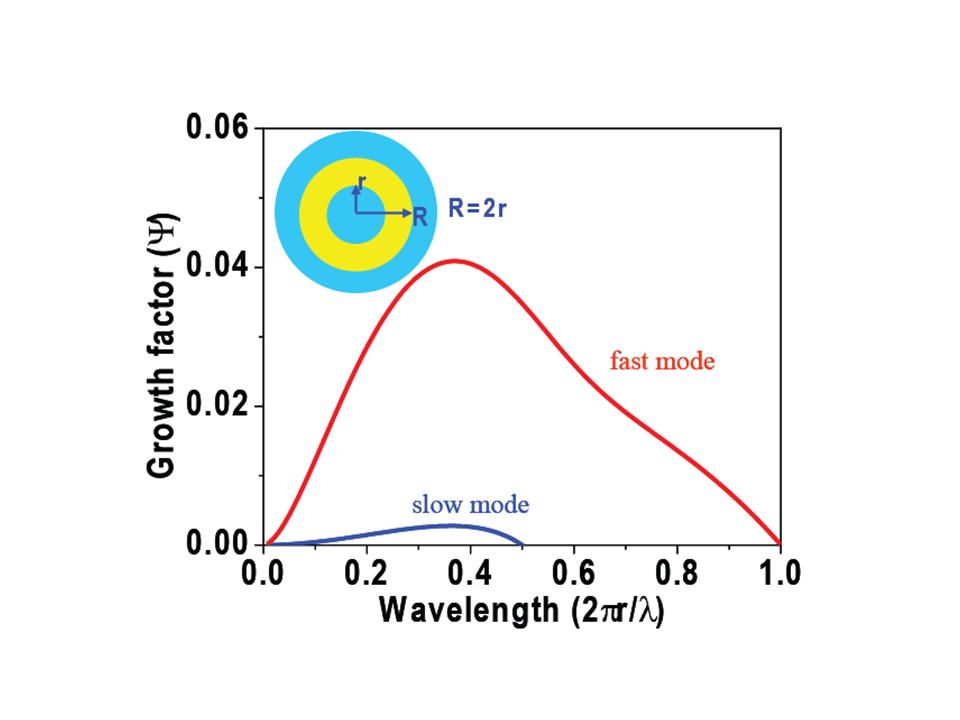}

\caption{\label{fig:growthfactor}Growth factor of instability as a function
of perturbation wavelength. Fast- and slow- modes occur at wavelengths
above their respective critical wavelengths $\lambda_{f},\lambda_{s}$.
Inset is a sketch of coaxial cylinder with radius $R=2r$ and equal
viscosities. }

\end{figure}

A linear theory of capillary instability for a co-axial cylinder with
equal viscosities is provided in the literature by Stone and Brenner
\cite{Stone1996}. The growth rate ($\sigma$) for a wave vector $k=2\pi/\lambda$
is a solution of the following quadratic equations, \begin{multline}
\left\{ \sigma-\frac{k^{2}\gamma_{1}}{r\eta}\left[1-(rk)^{2}\right]\Lambda(r,r)\right\} \\
\times\left\{ \sigma-\frac{k^{2}\gamma_{2}}{R\eta}\left[1-(Rk)^{2}\right]\Lambda(R,R)\right\} \\
=\frac{k^{4}\gamma_{1}\gamma_{2}}{rR\eta^{2}}\left[1-(rk)^{2}\right]\left[1-(Rk)^{2}\right]\Lambda(r,R)^{2},\label{eq:Analyticformula}\end{multline}
where $r$ and $R$ are the radii of the unperturbed interfaces~I
and~II, $\gamma_{1}$ and $\gamma_{2}$ are the interfacial tensions,
and $\eta$ is viscosity. $\Lambda(a,b)$, where $a\leq b$, is associated
with the modified Bessel function, 

\begin{equation}
\Lambda(a,b)=\int_{0}^{\infty}\frac{sJ_{1}(sa)J_{1}(sb)}{(s^{2}+k^{2})^{2}}ds=-\frac{1}{2k}\frac{d}{dk}\left[I_{1}(ak)K_{1}(bk)\right]\label{eq:Besselassociated}\end{equation}
For the case of $\gamma_{1}=\gamma_{2}=\gamma$, the growth rate has
the following formula,

\begin{equation}
\sigma(\lambda)=\frac{\gamma}{\eta r}\Psi(\lambda,R/r),\label{eq:Growthrate}\end{equation}
where the growth factor of $\Psi(\lambda,R/r)$ in Eq. \ref{eq:Growthrate}
is a complicated function of instability wavelength \cite{Stone1996}.
The instability time scale $\tau\sim\sigma^{-1}\sim\eta r/\gamma$
is scaled with radius. For the case of $R=2r$, this growth factor
is calculated in Fig. \ref{fig:growthfactor}. A positive growth factor
indicates a positive growth rate ($\sigma>0$), for which any perturbation
is exponentially amplified with time. Instability occurs at long wavelengths
above a certain critical wavelength. Two critical wavelengths exist
for the co-axial cylinder shell. One is a short critical wavelength
$\lambda_{f}=2\pi r$ for a faster-growth mode (red line). The other
is a long critical wavelength $\lambda_{s}=2\pi R$ for slower-growth
mode (blue line). In the numerical simulation, the wavelength is chosen
between these two wavelengths ($\lambda_{f}<\lambda<\lambda_{s}$)
, and fast modes dominate. From the simulation parameters $\eta\approx10^{5}$
$\mathrm{Pa\cdot s}$ and $\gamma/r\approx10^{4}$ $\mathrm{Pa\cdot s}$,
together with the wavelength $2\pi r/\lambda\approx0.47$ corresponding
to a growth factor $\Psi(\lambda)\approx0.03$, the linear theory
predicts an instability time scale $\tau=\sigma^{-1}=(\frac{\gamma}{\eta r}\Psi(\lambda,R/r))^{-1}\approx334$
$\mathrm{sec}$.

\section*{3. Viscous materials during thermal drawing }

\begin{figure}
\includegraphics[width=1\columnwidth]{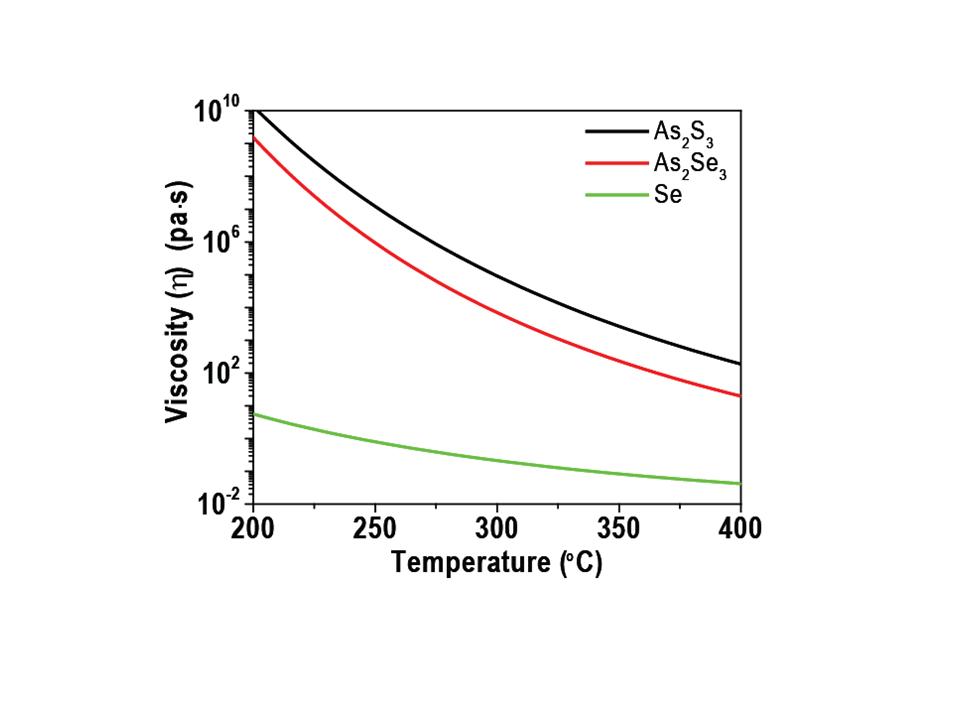}\caption{\label{fig:viscositytemp}Temperature-dependent viscosity for various
chalcogenide glasses. Typical temperature during fiber drawing for
glass $\mathrm{Se,As_{2}Se_{3},As_{2}S_{3}}$ is around $220,260,300$
\textcolor{black}{$\mathrm{^{o}C}$ with the corresponding viscosities
of $10,10^{5},10^{5}$ $\mathrm{Pa\cdot s}$, respectively}. }

\end{figure}

\textcolor{black}{Our chosen materials include chalcogenide glasses
($\mathrm{Se}$, }$\mathrm{As_{2}Se_{3}}$,$ $ \textcolor{black}{and
$\mathrm{As_{2}S_{3}}$) and thermoplastic polymers ($\mathrm{PES}$,
$\mathrm{PEI}$, and $\mathrm{PSU}$). }The viscosity of chalcogenide
glass-forming melts depends on temperature and is calculated from
an empirical Arrhenius formula\cite{Debenedetti2001}, 

\begin{equation}
\mathrm{log}\eta=\mathrm{log}\eta_{0}+C\frac{e^{D/T}}{2.3RT}-1,\label{eq:viscosity}\end{equation}
\textcolor{black}{where $R$ is the ideal gas constant, $T$ is the
temperature in Kelvin, and $\eta$ is viscosity in $\mathrm{Pa\cdot s}$.
The parameters of $\mathrm{log}\eta_{0}$, $C,$ and $D$ for our
materials are listed below: $-2.0,\mbox{}6651,\mbox{}770.82$ for
$\mathrm{Se}$, $-3.09,\mbox{}18877.8,\mbox{}875.56$ for $\mathrm{As_{2}Se_{3}}$,
and $-3.62,\mbox{}33744,\mbox{}650.8$ for $\mathrm{As_{2}S_{3}}$\cite{Tverjanovich2003}.
These viscosities over a wide temperature range are plotted in Fig.
\ref{fig:viscositytemp}. The typical temperature during a fiber drawing
for $\mathrm{Se,\mbox{}As_{2}Se_{3}}$, or }$\mathrm{As_{2}S_{3}}$\textcolor{black}{{}
films is around $220,260$, or $300$ $\mathrm{^{o}C}$, respectively,
with the corresponding viscosities of $10,10^{5}$, or }$10^{5}$\textcolor{black}{{}
$\mathrm{Pa\cdot s}$, respectively. }

\section*{4. Diffusion term in level-set equation}

To ensure numerical stability in the simulation, an artificial diffusion
term proportional to a small parameter $D$ is added to level-set
Eq. \ref{eq:Levelset} as follows $ $\textcolor{black}{{} \cite{Olsson2005}, }

\[
\phi_{t}+\vec{u}\cdot\vec{\nabla}\phi=D\nabla\cdot\left\{ \left[1-\phi(1-\phi)\right]\nabla\phi\right\} .\]

\end{document}